\documentclass[twocolumn]{aastex63}

\usepackage{amssymb,amsmath,graphicx,latexsym}
\usepackage{xcolor}[dvipsnames]

\usepackage{subfigure}

\usepackage{multirow}

\received{???}
\revised{???}
\accepted{???}

\submitjournal{ApJ}

\shorttitle{Cascading Dark Energy}
\shortauthors{Rezazadeh et al.}

\begin{document}

\title{Cascading Dark Energy}

\correspondingauthor{Kazem Rezazadeh}
\email{kazem.rezazadeh@ipm.ir}

\author{Kazem Rezazadeh}
\affiliation{School of Physics, Institute for Research in Fundamental Sciences (IPM),
P.O. Box 19395-5531, Tehran, Iran}

\author{Amjad Ashoorioon}
\affiliation{School of Physics, Institute for Research in Fundamental Sciences (IPM),
P.O. Box 19395-5531, Tehran, Iran}

\author{Daniel Grin}
\affiliation{Department of Physics and Astronomy, Haverford College, 370 Lancaster Avenue, Haverford, PA 19041, United States}



\begin{abstract}

The standard cosmological model is in the midst of a stress test, thanks to the tension between supernovae-based measurements of the Hubble constant $H_{0}$ and inferences of its values from Cosmic Microwave Background (CMB) anisotropies. Numerous explanations for the present-day cosmic acceleration require the presence of a new fundamental scalar field, as do Early Dark Energy (EDE) solutions to the Hubble tension. This raises the possibility that multiple fields cooperatively contribute to the dark energy component in bursts throughout cosmic time due to distinct initial conditions and couplings. Here, this Cascading Dark Energy (CDE) scenario is illustrated through a realization that effectively reduces to a two-field model, with two epochs in which dark energy is cosmologically significant. The model is compared to measurements of the CMB, baryon acoustic oscillations, as well as both PANTHEON and SH0ES observations of Type-Ia supernovae. Neglecting the linear perturbations, it is found that this scenario ameliorates the Hubble tension, improving over purely late-time models of dark energy, and the agreement between the galaxy survey measurements of baryon acoustic oscillations.

\end{abstract}




\section{Introduction}
\label{section:introduction}

Measurements of the present-day Hubble parameter (the Hubble constant $H_{0}$) from supernovae \citep{Riess:2016jrr, Riess:2018byc, Riess:2019cxk, Riess:2021jrx, Riess:2022mme, Murakami:2023xuy} disagree with the value inferred from the Cosmic Microwave Background (CMB) data \citep{Planck:2018vyg, Planck:2019nip, Planck:2018lbu} fit the $\Lambda$CDM model. More specifically, CMB power spectra determined by the \textit{Planck} collaboration yield a preferred value of $H_{0}=67.4\pm0.6\,{\rm km\,s^{-1}\,Mpc^{-1}}$ \citep{Planck:2018vyg}. Similarly, measurements of the acoustic horizon by the Dark Energy Survey (DES), combined with constraints to the baryon density from Big-Bang Nucleosynthesis (BBN) abundances, yield the value $H_{0}=67.4_{-1.2}^{+1.1}\,{\rm km\,s^{-1}\,Mpc^{-1}}$ \citep{DES:2017txv}. Strong-lensing time delays also probe $H_{0}$, but realistic error bars (which include the mass-sheet degeneracy) leave these measurements consistent with both CMB and supernovae values.\footnote{Strong-lensing time delays also probe $H_{0}$ \citep{Bonvin:2016crt, Birrer:2018vtm}, but realistic error bars (which include the mass-sheet degeneracy) leave these measurements consistent with both CMB and supernovae values \citep{Millon:2019slk, Shajib:2023uig}.}

In contrast, observations of Type-Ia supernovae, tethered to a distance ladder obtained using Hubble Space Telescope (HST) measurements of $70$ long-period Cepheids in the Large Magellanic Cloud, imply a substantially different value, $H_{0}=73.3\pm 1.1 \,\mathrm{km\ s^{-1}Mpc^{-1}}$\citep{Riess:2022mme}. This is known as the ``Hubble tension''. Although this discrepancy might be caused by systematic effects in the data (though none seem sufficient so far), it could alternatively herald exciting new physics beyond the $\Lambda$CDM concordance model.

Many proposals have been suggested to resolve this tension [see, e.g. \citep{Umilta:2015cta, Karwal:2016vyq, Poulin:2018dzj, Poulin:2018cxd, Pandey:2019plg, Agrawal:2019lmo, Vagnozzi:2019ezj, Smith:2019ihp, Davari:2019tni, Sola:2020lba, Smith:2020rxx, Krishnan:2020vaf}], some of which (late-time resolutions) invoke modifications to $\Lambda$CDM which become predominant near the current cosmological epoch, others of which cause modifications to the cosmic budget at early times (early-time resolutions, which modify cosmic evolution around or before matter-radiation equality).

Among the proposed late-time resolutions are phantom-like Dark Energy (DE) \citep{DiValentino:2016hlg, DiValentino:2017zyq}, a vacuum phase transition \citep{DiValentino:2017rcr}, interacting DE \citep{Kumar:2016zpg, DiValentino:2017iww}, and modified theories of gravity \citep{Barreira:2014jha, Umilta:2015cta, Ballardini:2016cvy, Renk:2017rzu, Belgacem:2017cqo, Nunes:2018xbm, Lin:2018nxe}. These scenarios \citep{DiValentino:2017zyq, DiValentino:2017iww, Addison:2017fdm} and more model-independent generalizations of them \citep{Bernal:2016gxb, Zhao:2017cud, Poulin:2018zxs} are highly constrained by the data, especially by measurements of the Baryon Acoustic Oscillations (BAO) \citep{Beutler:2011hx, Ross:2014qpa, BOSS:2016wmc} in galaxy surveys. Late-time resolutions to the Hubble tension usually suffer from some fundamental shortcomings, such as being a worse fit to CMB data than $\Lambda$CDM, fine-tuning issues, inappropriate use of an $H_{0}$ prior \citep{Efstathiou:2021ocp}, and conflicts with the ages of globular clusters \citep{Jimenez:2019onw, Valcin:2020vav, Valcin:2021jcg}, as discussed in \citet{Bernal:2021yli}.

In most realizations of early-time solutions, the sound horizon is reduced by introducing additional radiation energy density to the matter-energy content of the Universe. Such scenarios are also constrained by BAO and the high-$\ell$ CMB power spectrum \citep{Poulin:2018cxd, Smith:2019ihp, Karwal:2021vpk, Poulin:2021bjr, Murgia:2020ryi}. For example, in Early Dark Energy (EDE) scenarios, the Universe contains a component (typically a scalar field) whose behavior is like a cosmological constant prior to a critical redshift (preceding matter-radiation equality) and dilutes as fast or faster than radiation \citep{Poulin:2018cxd, Smith:2019ihp} subsequently.

Aside from the Hubble tension, there is another disagreement between cosmological data sets, known as the $S_8$ tension. This tension reflects the fact that the value of $S_{8}\equiv\sigma_{8}\sqrt{\Omega_{\rm m}/0.3}$ (where $\Omega_{\rm m}$ is the today's matter density and $\sigma_8$ denotes the variance of matter perturbations within $8\mathrm{Mpc}/h$ today) implied by the CMB (when fit by the $\Lambda$CDM model) does not agree with the value inferred from galaxy weak-lensing measurements of the amplitude of matter density fluctuations in the late-time Universe \citep{DES:2017myr, Hildebrandt:2018yau, HSC:2018mrq, KiDS:2020suj, KiDS:2021opn, DES:2021wwk, DES:2021epj}.

Results from Dark Energy Survey (DES) 3-year data yield the constraint $S_{8}=0.797_{-0.013}^{+0.015}$ (68\% CL) \citep{DES:2021epj}, in contrast with the value $S_{8}=0.832\pm0.013$ (68\% CL) implied by the best-fit value for the amplitude of the scalar density power spectrum from Planck 2018 TT,TE,EE +lowE+CMB lensing data, assuming ${\Lambda}$CDM \citep{Planck:2018vyg}. Although EDE models can alleviate the $H_0$ tension, they tend to exacerbate the $S_8$ tension \citep{Hill:2020osr, Ivanov:2020ril, Murgia:2020ryi} and worsen fits to BAO data [see, e.g. \citep{Poulin:2018cxd, Smith:2019ihp, Murgia:2020ryi}].

It is interesting to consider the possibility that these tensions (and the required fine-tuning of EDE models) could be alleviated by a richer dark energy sector, for example, if there are multiple epochs of cosmic acceleration driven by one field \citep{Niedermann:2019olb, Niedermann:2021vgd, Freese:2021rjq, Allali:2021azp}, or if many scalar fields acting over time could yield better concordance between cosmological data sets \citep{Sabla:2021nfy, Ramadan:2023ivw}.

In this paper, we explore if a resolution for the Hubble tension can be found in the Cascading Dark Energy (CDE) scenario, in which multiple scalar fields contribute to dark energy, analogously to the assisted inflationary scenario \citep{Liddle:1998jc}. CDE is motivated by recent developments in string theory, such as the swampland conjecture \citep{Vafa:2005ui, Ooguri:2006in, Obied:2018sgi}. CDE reduces the Hubble tension primarily by altering the early-time sound horizon, like the standard EDE scenario.
 
Some of the fields, however, drop out of sync with the others due to their initial conditions - that is to say that they no longer roll slowly (with nearly constant energy density) even as other fields jointly continue to behave as dark energy. The evolution of each field becomes significant after the Hubble parameters drop below some specific value which depends on the effective mass of the field as well as the background energy density. After that, the field begins to oscillate around the local minimum of its potential and loses its energy accordingly.

In the simplest realization, our model will reduce to two fields, allowing us to treat the dynamics of each field separately without resorting to an effective one-field approximation, as was done in, e.g. \citet{Sabla:2021nfy}. Multi-field models for dark energy are well motivated by considerations from string theory, such as the axiverse scenario \citep{Arvanitaki:2009fg}, in which a broad mass spectrum of ultra-light axions could contribute to both dark matter and dark energy \citep{Hlozek:2014lca, Marsh:2015xka}. They may contribute to explaining the ``why now" question for the late-time dark energy driving present-day cosmic acceleration \citep{Kamionkowski:2014zda, Emami:2016mrt}. At earlier times, multiple-field scenarios could help reduce the fine-tuning needed for EDE models to succeed. Here, we consider the possibility that some dark energy fields are relevant near equality/recombination, while others are more relevant today.

We investigate the behavior of the Hubble parameter and the field configuration in our setup. Both fields couple to gravity minimally, and their kinetic terms are assumed to be canonical. The potentials of both fields in the simplest realization are considered to be quartic, although one can assume that they are different, as we will explain later. In our work, we check the consistency of the CDE scenario with the existing data, including the CMB \citep{Planck:2018vyg, Planck:2019nip, Planck:2018lbu}, Pantheon SN \citep{Pan-STARRS1:2017jku}, BAO \citep{BOSS:2016wmc, Ross:2014qpa, Beutler:2012px}, and \cite{Riess:2019cxk} (SH0ES) measurements. We use Monte Carlo Markov Chains (MCMC) simulations to constrain CDE model parameters using cosmological data. We compare the CDE scenario with the concordance $\Lambda$CDM model, as well as with a single-field canonical scalar field that couples minimally to gravity and has a quartic potential.

Furthermore, we compare our model to the Rock `n' Roll (RnR) quartic model \citep{Agrawal:2019lmo}, which includes a cosmological constant and an oscillating scalar field acting as EDE. It is known that in the case of a full $(1-\cos{\theta})^{n}$ potential [as considered in \citet{Poulin:2018cxd, Smith:2019ihp}], anharmonic deviations from quadratic behavior are important in driving perturbative mode evolution towards behavior that more optimally addresses the $H_{0}$ tension than the Rock `n' Roll scenario. Nevertheless, we compare our CDE scenario to the Rock `n' Roll realization of EDE, as it provides a useful foil for comparing single- and multi-field models with similar potentials.

We compare the result of our two-field CDE scenario for $H_0$ with the results of the $\Lambda$CDM model, the single-field DE model, and also the Rock `n' Roll scenario. We want to know if the $H_0$ tension can be resolved via the CDE framework, and if so, whether it has any advantages. The full dynamics of \emph{two} coherently oscillating scalar fields coupled through gravity can be complicated and potentially computationally expensive [e.g. \citet{Chen:2023unc}], especially in the presence of perturbations. A complete analysis with minimal priors would ultimately require a careful analysis of resonance between the two fields (both for perturbations and the background), as well as multi-field initial conditions for the perturbations. In this work, our primary aim is to assess the relative merits of CDE compared with the Rock `n' Roll model and to see if the addition of only one additional field can significantly impact EDE's resolution of the Hubble tension. We comment more on these simplifications later in the paper.

We compare the implied value of $S_{8}$ for the empirically allowed parameter space of our model with that of these other models. Altogether, we find that the two-field CDE model ($\chi_{\mathrm{total}}^{2}=3828.02$) fits the observational data better than the $\Lambda$CDM ($\chi_{\mathrm{total}}^{2}=3832.03$) and single-field DE ($\chi_{\mathrm{total}}^{2}=3832.17$) models. The result of our two-field CDE model is more consistent with the Riess et al.~(2019) measurement in comparison with the predictions of the $\Lambda$CDM and single-field DE models, and therefore our model can ameliorate the $H_0$ tension that exists between cosmological data from different sources. Due to the resemblance of our two-field model to the Rock `n' Roll model with $n=2$, we contrast our setup with that model too. We find that the predictions of quartic two-field CDE are very close to the Rock `n' Roll model with $n=2$, although the late-time evolution of the dark energy at late times in our model yields a modestly better fit to the BAO data. Overall, however the Rock `n' Roll model still fits the data sets better due to the worse fit of CDE to the SH0ES data.

The rest of this paper is structured as follows: In Sec.~\ref{section:motivation}, we introduce the CDE model and explain its theoretical motivation. Then, in Sec.~\ref{section:setup}, we explore the two-field realization of the CDE model and present its equations of motion. Subsequently, in Sec.~\ref{section:numerical}, we describe the setup of a Monte Carlo Markov Chain (MCMC) analysis used to test a two-field CDE scenario using cosmological data. We discuss results in Sec. \ref{section:results}. We present our conclusions in Sec. \ref{section:conclusions}, where we also put forward avenues to expand and further test the CDE scenario.


\section{Setup of Cascading Dark Energy}
\label{section:motivation}

We consider $N+1$ scalar fields with quartic monomial potentials, with $N \gg 1$, $V(\phi_i)=\frac{\lambda}{4} \phi_i^4$, $i=1\ldots N+1$, with the Lagrangian
\begin{equation}\label{Lag1}
S=\int \sqrt{-g} d^4 x \left[\sum_{i=1}^{N+1}\left(\frac{1}{2}\partial_{\mu} \phi_i \partial^{\mu} \phi_i-\frac{\lambda}{4} {\phi}_i^4\right) \right] \, .
\end{equation}
For simplicity, we have assumed that the quartic couplings of all the fields are the same. In principle, these scalar fields can have different initial conditions. We assume the swampland distance conjecture, under which these fields can at most transverse $M_P$ in the field space before a tower of massless species appears. We thus assume that the initial conditions of all the fields are sub-Planckian. Following the de-Sitter swampland conjecture, we also assume that the relative slope of the potential should not be very flat, yielding the constraint that
\begin{equation}\label{desitter-swampland}
M_P\frac{V'}{V}\gtrsim c=\mathcal{O}(1)\,.
\end{equation}
Let us assume that all the first $N$ fields have the same initial conditions, which are different from that of the $(N+1)$-th field,
\begin{eqnarray}
\label{initcond}
&&\phi_1=\phi_2=\ldots=\phi_N=\phi_0\,,\nonumber  \\
&&\phi_{N+1}= \chi_0\,. 
\end{eqnarray}
Then the effective Lagrangian of $\phi_0$ and $\chi_0$ can be written as,
\begin{align}
S= & \int d^{4}x\sqrt{-g}\bigg(\frac{N}{2}\partial_{\mu}\phi_{0}\partial^{\mu}\phi_{0}+\frac{1}{2}\partial_{\mu}\chi_{0}\partial^{\mu}\chi_{0}
\nonumber
\\
& -N\frac{\lambda}{4}\phi_{0}^{4}-\frac{\lambda}{4}\chi^{4}\bigg) \,.
\label{Lag2}
\end{align}
We introduce the new effective fields, $\phi$ and $\chi$,
\begin{align}
\phi &\equiv \sqrt{N} \phi_0\,,\nonumber\\
\chi &\equiv \chi_0\,,
\label{field-dressing}
\end{align}
to make the kinetic term of the $\phi_0$ field in the Lagrangian canonical, which leads to the Lagrangian
\begin{equation}
\label{Lag3}
S = \int\sqrt{-g}d^{4}x\bigg(\frac{1}{2}\partial_{\mu}\phi\partial^{\mu}\phi+\frac{1}{2}\partial_{\mu}\chi\partial^{\mu}\chi -\frac{\lambda_{\phi}}{4}\phi^{4}-\frac{\lambda_{\chi}}{4}\chi^{4}\bigg) \, ,
\end{equation}
where
\begin{eqnarray}\label{field-dressing2}
\lambda_{\phi}&\equiv& \frac{\lambda}{N},\nonumber\\
\lambda_{\chi}&\equiv& \lambda\,,
\end{eqnarray}
Although the fields $\phi_i$, $i=1\ldots N+1$, cannot be super-Planckian due to swampland conjecture \citep{Ooguri:2006in}, the fields $\phi$ can be super-Planckian due to the large dressing factor $\sqrt{N}$ if $N\gg 1$.

With these initial conditions, the fields can act as dark energy components in our setup. The other notable thing is that if $N \gg 1$, the quartic couplings of the $\chi$ field become much larger than the $\phi$ field, whereas the initial condition for the $\chi$ field becomes smaller and, in fact, sub-Planckian compatible with the swampland conjecture. Due to these, the $\chi$ field can play the role of a cascade field in our setup, which starts to oscillate around its minimum after the Hubble parameter squared drops below its mass, $\partial_{\chi}^2 V(\chi)$, and its energy density becomes a substantial part of the background energy density. This will lead to a sudden drop of the comoving sound horizon before the decoupling, which enhances the Hubble parameter respectively today if the angular $\theta_{\rm MC}$ parameter is fixed by the CMB experiments.

A similar model could be constructed in the context of multi-giant matrix vacua \citep{Ashoorioon:2009sr, Ashoorioon:2009wa, Ashoorioon:2011ki, Ashoorioon:2014jja} that uses concentric multiple stacks of D3-branes. In that model, the matrix structure of the coordinates perpendicular to the stack of D3-branes, and the ansatz of the SU(2) generator for three of the orthogonal directions perpendicular to the stacks of D3-branes is used. To exploit the model to describe the late time Universe, with the string coupling $g_{{}_S}\sim 1$, one has to use a large number of D3-branes, $N\sim 10^{40}$, and then one has to worry about to backreaction effects of the D3-branes on the background geometry. Alternatively, one can assume that $g_{{}_S}$ itself is extremely small, say $g_{{}_S}\sim 10^{-100}$ and further suppression of the quartic couplings of the fields to the required value to explain the dark energy vacuum density, $\lambda_{\phi,\chi}\sim 10^{-120}$ is achieved via the multiplicity of the D3-branes. Recently in \citet{Ashoorioon:2019kcy}, some of us showed that using a scalar field non-minimally coupled to gravity, with a moderate value of non-minimal coupling, one can reduce the required number of D3-branes to achieve the required number of D3-branes to a reasonable number during inflation. We will explore this scenario further in future work. Here, we instead focus on two minimally coupled scalar fields where their couplings, $\lambda$, are already small.


\section{The Two-Field Setup}
\label{section:setup}

In this work, we focus on the two-field realization of the CDE model, which consists of two dynamical scalar fields with canonical kinetic terms. The first Friedmann equation for a flat FRW Universe in this setup takes the following form
\begin{equation}
\label{H}
H^{2}=\frac{1}{3M_{P}^{2}}\left(\rho_{\mathrm{\rm m}}+\rho_{\mathrm{\rm r}}+\rho_{\phi}+\rho_{\chi}\right),
\end{equation}
where $H=\dot{a}/a$ is the Hubble parameter and $M_{\rm P}\equiv1/\sqrt{8\pi G}$ is the reduced Planck mass. Furthermore, $\rho_{\rm m}$ and $\rho_{\rm r}$ denote the energy densities of matter and radiation, respectively. The energy densities of the scalar field $\phi$ and $\chi$ are respectively denoted by $\rho_{\phi}$ and $\rho_{\chi}$. It should be noted that in our work, we assume that the neutrinos are massless, and therefore their contribution is included in the energy density of the radiation component. The more general treatment of this setup requires also the inclusion of the massive neutrinos, and this possibility may be taken into account in future extensions of our scenario. It should also be noted that no definitive value for the mass of neutrinos has been reported so far.

The energy densities of matter and radiation vary with scale factor as follows
\begin{align}
\label{rhom}
\rho_{\rm m} &= \rho_{{\rm m}i}\left(\frac{a_{i}}{a}\right)^{3},\\
\label{rhor}
\rho_{\rm r} &= \rho_{{\rm r}i}\left(\frac{a_{i}}{a}\right)^{4},
\end{align}
where $\rho_{{\rm m} i}$ and $\rho_{{\rm r}i}$ are the energy densities of matter and radiation, respectively, at the initial scale factor $a_i$ that we take it deep inside in the radiation dominated era. We normalize these quantities as follows
\begin{align}
\label{rhomti}
\tilde{\rho}_{{\rm m}i} &\equiv \frac{\rho_{{\rm m}i}}{M_{\rm P}^{2}H_{0}^{2}},\\
\label{rhorti}
\tilde{\rho}_{{\rm r}i} &\equiv \frac{\rho_{{\rm r}i}}{M_{\rm P}^{2}H_{0}^{2}}\,,
\end{align}
where $H_0$ is the Hubble parameter today. We express the scale factor of the Universe in terms of the number of $e$-folds as
\begin{equation}
\label{a}
a=a_{i}e^{N}\,,
\end{equation}
and hence from Eqs. \eqref{rhom} and \eqref{rhor}, we find
\begin{align}
\label{rhom-N}
\rho_{\rm m} &= M_{\rm P}^{2}H_{0}^{2}\tilde{\rho}_{{\rm m}i}e^{-3N},\\
\label{rhor-N}
\rho_{\rm r} &= M_{\rm P}^{2}H_{0}^{2}\tilde{\rho}_{{\rm r}i}e^{-4N}.
\end{align}
The energy densities of the two canonical scalar fields are given by
\begin{align}
\label{rhophi}
\rho_{\phi} &= \frac{1}{2}\dot{\phi}^{2}+V_{\phi}(\phi),\\
\label{rhochi}
\rho_{\chi} &= \frac{1}{2}\dot{\chi}^{2}+V_{\chi}(\chi).
\end{align}
We take the potential of both the scalar fields in the following quartic forms
\begin{align}
\label{Vphi}
V_{\phi}(\phi) &= \frac{1}{4}\lambda_{\phi}\phi^{4},\\
\label{Vchi}
V_{\chi}(\chi) &= \frac{1}{4}\lambda_{\chi}\chi^{4},
\end{align}
where $\lambda_{\phi}$ and $\lambda_{\chi}$ are respectively the self-interaction coupling constants for the $\phi$ and $\chi$ fields. Applying the continuity equations for the energy densities Eqs. \eqref{rhophi} and \eqref{rhochi}, we obtain the equations of motion for $\phi$ and $\chi$, respectively, as
\begin{align}
\label{phiddot}
\ddot{\phi}+3H\dot{\phi}+\frac{dV_{\phi}(\phi)}{d \phi} &= 0\,,\\
\label{chiddot}
\ddot{\chi}+3H\dot{\chi}+\frac{dV_{\chi}(\chi)}{d \chi} &= 0\,.
\end{align}
Now, following  \citet{Rezazadeh:2020zrd}, we introduce the following normalized quantities
\begin{align}
& \tilde{H}\equiv\frac{H}{H_{0}},\qquad\tilde{\phi}\equiv\frac{\phi}{M_{P}},\qquad\tilde{\chi}\equiv\frac{\chi}{M_{P}},
\nonumber
\\
& \tilde{\lambda}_{\phi}\equiv\frac{M_{{\rm P}}^{2}}{H_{0}^{2}}\lambda_{\phi},\qquad\tilde{\lambda}_{\chi}\equiv\frac{M_{{\rm P}}^{2}}{H_{0}^{2}}\lambda_{\chi} \, .
\label{normalization}
\end{align}
As a result, from Eq. \eqref{H}, we get
\begin{equation}
\label{Ht}
\tilde{H}^{2}=\frac{4\left(\tilde{\rho}_{{\rm m}i}e^{-3N}+\tilde{\rho}_{{\rm r}i}e^{-4N}\right)+\tilde{\lambda}_{\phi}\tilde{\phi}^{4}+\text{\ensuremath{\tilde{\lambda}_{\chi}}}\tilde{\chi}^{4}}{2\left(6-\tilde{\phi}'^{2}-\tilde{\chi}'^{2}\right)},
\end{equation}
where the prime denotes the derivative with respect to the $e$-fold number $N$. If we take the derivative of both sides of the above equation with respect to $N$, we obtain
\begin{align}
\tilde{H}'= & -\frac{1}{\tilde{H}\left(6-\tilde{\phi}'^{2}-\tilde{\chi}'^{2}\right)}\bigg[3\tilde{\rho}_{{\rm m}i}e^{-3N}+\tilde{\rho}_{{\rm r}i}4e^{-4N}
\nonumber
\\
& -\tilde{H}^{2}\left(\tilde{\phi}'\tilde{\phi}''+\tilde{\chi}'\tilde{\chi}''\right)-\tilde{\lambda}_{\phi}\tilde{\phi}^{3}\tilde{\phi}'-\tilde{\lambda}_{\chi}\tilde{\chi}^{3}\tilde{\chi}'\bigg] \, .
\label{dHt-d2phit-d2chit}
\end{align}
Applying the normalized quantities \eqref{normalization} in \eqref{phiddot} and \eqref{chiddot}, we also reach
\begin{align}
\label{d2phit-dHt}
\tilde{\phi}'' &= -\frac{\tilde{H}\left(3\tilde{H}+\tilde{H}'\right)\tilde{\phi}'+\tilde{\lambda}_{\phi}\tilde{\phi}^{3}}{\tilde{H}^{2}}\,,\\
\label{d2chit-dHt}
\tilde{\chi}'' &= -\frac{\tilde{H}\left(3\tilde{H}+\tilde{H}'\right)\tilde{\chi}'+\tilde{\lambda}_{\chi}\tilde{\chi}^{3}}{\tilde{H}^{2}}\,.
\end{align}
Inserting these into Eq. \eqref{dHt-d2phit-d2chit}, and then solving the resulting equation for $\tilde{H}'$, we arrive at
\begin{equation}
\label{dHt}
\tilde{H}'=-\frac{1}{6\tilde{H}e^{4N}}\left[3\tilde{H}^{2}\left(\tilde{\phi}'^{2}+\tilde{\chi}'^{2}\right)e^{4N}+3\tilde{\rho}_{{\rm m}i}e^{N}+4\tilde{\rho}_{{\rm r}i}\right].
\end{equation}
To eliminate $\tilde{H}'$ in Eqs. \eqref{d2phit-dHt} and \eqref{d2chit-dHt}, we use the above equation and hence we will have
\begin{align}
\tilde{\phi}''= & -\frac{1}{6\tilde{H}^{2}}\bigg[e^{-4N}\text{\ensuremath{\tilde{\phi}}}'\bigg(3e^{4N}\tilde{H}^{2}\left(6-\tilde{\chi}'^{2}-\tilde{\phi}'^{2}\right)
\nonumber
\\
& -3\tilde{\rho}_{{\rm m}i}e^{N}-4\tilde{\rho}_{{\rm r}i}\bigg)+6\tilde{\lambda}_{\phi}\tilde{\phi}^{3}\bigg] \, ,
\label{d2phit}
\\
\tilde{\chi}''= & -\frac{1}{6\tilde{H}^{2}}\bigg[e^{-4N}\text{\ensuremath{\tilde{\chi}}}'\bigg(3e^{4N}\tilde{H}^{2}\left(6-\tilde{\chi}'^{2}-\tilde{\phi}'^{2}\right)
\nonumber
\\
& -3\tilde{\rho}_{{\rm m}i}e^{N}-4\tilde{\rho}_{{\rm r}i}\bigg)+6\tilde{\lambda}_{\chi}\tilde{\chi}^{3}\bigg] \, .
\label{d2chit}
\end{align}
Eqs. \eqref{dHt}, \eqref{d2phit}, and \eqref{d2chit}, are basic equations that we will solve in our work to find the background dynamics. To determine the initial conditions, we assume that the evolution of the two scalar fields starts from the slow-roll regime. Therefore, the first term in Eqs. \eqref{phiddot} and \eqref{chiddot} can be neglected before the other terms, and these equations simplify as follows
\begin{align}
\label{phidot}
3H\dot{\phi}+\frac{dV_{\phi}(\phi)}{d\phi} &\approx 0,\\
\label{chidot}
3H\dot{\chi}+\frac{dV_{\chi}(\chi)}{d\chi} &\approx 0\,,
\end{align}
which in turn can be written in terms of $e$-folds number as
\begin{align}
\label{dphit}
\tilde{\phi}' &\approx -\frac{\tilde{\lambda}_{\phi}\tilde{\phi}^{3}}{3\tilde{H}^{2}},\\
\label{dchit}
\tilde{\chi}' &\approx -\frac{\tilde{\lambda}_{\chi}\tilde{\chi}^{3}}{3\tilde{H}^{2}}.
\end{align}
Besides, in the slow-roll regime, the kinetic terms of $\phi$ and $\chi$ are negligible in comparison with their potentials, and therefore the Friedmann equation, Eq.~\eqref{H}, can be approximated as
\begin{equation}
\label{H-sr}
H^{2}\approx\frac{1}{3M_{P}^{2}}\left[\rho_{\rm m}+\rho_{\rm r}+V_{\phi}(\phi)+V_{\chi}(\chi)\right].
\end{equation}
If we substitute $\rho_{\rm m}$ and $\rho_{\rm r}$ from Eqs. \eqref{rhom} and \eqref{rhor}, respectively, into this equation, and then use Eqs. \eqref{rhom-N}, \eqref{rhor-N}, \eqref{Vphi}, \eqref{Vchi}, and \eqref{normalization}, we reach
\begin{equation}
\label{Ht-sr}
\tilde{H}^{2}\approx\frac{1}{12}\left[4\left(\tilde{\rho}_{{\rm m}i}e^{-3N}+\tilde{\rho}_{\mathrm{ r} i}e^{-4N}\right)+\tilde{\lambda}_{\phi}\tilde{\phi}^{4}+\tilde{\lambda}_{\chi}\tilde{\chi}^{4}\right].
\end{equation}
This equation now can be inserted into Eqs. \eqref{dphit} and \eqref{dchit} to give the initial values of derivative of the two scalar fields with respect to $N$ as
\begin{align}
\label{dphiti}
\tilde{\phi}_{i}' &\approx -\frac{4\tilde{\lambda}_{\phi}\tilde{\phi}_{i}^{3}}{4\left(\tilde{\rho}_{{\rm m}i}+\tilde{\rho}_{{\rm r}i}\right)+\tilde{\lambda}_{\phi}\tilde{\phi}_{i}^{4}+\tilde{\lambda}_{\chi}\tilde{\chi}_{i}^{4}},\\
\label{dchiti}
\tilde{\chi}_{i}' &\approx -\frac{4\tilde{\lambda}_{\chi}\tilde{\chi}_{i}^{3}}{4\left(\tilde{\rho}_{{\rm m}i}+\tilde{\rho}_{{\rm r}i}\right)+\tilde{\lambda}_{\phi}\tilde{\phi}_{i}^{4}+\tilde{\lambda}_{\chi}\tilde{\chi}_{i}^{4}}.
\end{align}
In these equations, $\tilde{\phi}_{i}$ and $\tilde{\chi}_{i}$ refer to the initial values of the scalar fields at the $e$-fold number $N_i = 0$. Because $H$, $\phi$, and $\chi$ are all numerically evolved in our representation of the dynamics, accurate initial conditions are needed for all these quantities to avoid exciting undesirable numerical transients. It should be noted that we cannot use Eq.~\eqref{Ht-sr} as the initial condition for $\tilde{H}$, because it leads to self-inconsistency of the differential equations. Instead, to prevent this problem, we apply the following relation which follows from Eq.~\eqref{Ht},
\begin{equation}
\label{Hti}
\tilde{H}_{i}^{2} = \frac{4\left(\tilde{\rho}_{{\rm m}i}+\tilde{\rho}_{{\rm r}i}\right)+\tilde{\lambda}_{\phi}\tilde{\phi}_{i}^{4}+\tilde{\lambda}_{\chi}\tilde{\chi}_{i}^{4}}{2\left(6-\tilde{\phi}_{i}'^{2}-\tilde{\chi}_{i}'^{2}\right)}.
\end{equation}
For $\tilde{\phi}'$ and $\tilde{\chi}'$ in this equation, we substitute their values from the slow-roll equations \eqref{dphiti} and \eqref{dchiti}, respectively.

To integrate the background equations \eqref{dHt}, \eqref{d2phit}, and \eqref{d2chit}, we used the 8th-order Runge-Kutta algorithm. Our modified version of \textsc{Camb}, (which we use with CosmoMC to obtain constraints on the CDE model), is available online\footnote{https://github.com/krezazadeh/CAMB-CDE-two-field}. In order to ensure that the Universe always remains flat in our code for each set of input parameters, we use the parameter $\tilde{\lambda}_\phi$ as a derived parameter. To determine this parameter numerically, we note that the total density parameter at the present epoch is equal to the unity for a flat Universe,
\begin{equation}
 \label{Omegatotal}
 \Omega_{\rm m0}+\Omega_{\rm r0}+\Omega_{\phi0}+\Omega_{\chi0}=1,
\end{equation}
where the subscript ``0'' refers to the present time. From this equation, we find
\begin{equation}
\label{lambdaphit}
\tilde{\lambda}_{\phi}=\frac{12-\tilde{\lambda}_{\chi}\tilde{\chi}_{0}^{4}-2\tilde{\phi}_{0}'^{2}-2\tilde{\chi}_{0}'^{2}-12\Omega_{m0}-12\Omega_{r0}}{\tilde{\phi}_{0}^{4}}.
\end{equation}
We use a shooting method in our numerical code that tests different values for $\tilde{\lambda}_\phi$ in the above equation for each set of free parameters. After several steps, the code finally finds a suitable value for this parameter that satisfies this equation with enough precision. As a result the parameter $\tilde{\lambda}_\phi$ is treated as a derived parameter in our numerical analysis. By requiring the $\phi$ field to provide the energy density required for a flat universe today, we target scenarios in which the data require $\chi$ to act as an EDE field and $\phi$ to be the present-day DE.


\section{Numerical analysis}
\label{section:numerical}

Here, we obtain observational constraints to the CDE model at the level of background dynamics, using recent cosmological data. We use the publicly available CosmoMC computational package \citep{Lewis:2002ah}. In this work, we used the July 2019 version of CosmoMC. This code uses a Markov Chain Monte Carlo (MCMC) simulation to explore the parameter space of the model, using the Metropolis-Hastings algorithm \citep{Lewis:2002ah}.

Our parameter space consists of \{$\Omega_b h^2$, $\Omega_c h^2$, $\theta_{\rm MC}$, $\tau$, $A_s$, $n_s$, $\tilde{\phi}_i$, $\tilde{\chi}_i$, $\tilde{\lambda}_{\chi}$\}, where $\Omega_b$ and $\Omega_c$ denote the present-day density parameters for baryon and cold dark matter, $h$ is the dimensionless Hubble constant $h\equiv H_{0}/(100~{\rm km}{\rm s}^{-1}~{\rm Mpc}^{-1})$, $\theta_{\rm MC}$ refers to the ratio of the comoving sound horizon at decoupling to the comoving angular diameter distance to the surface of last scattering, $\tau$ indicates the optical depth, $A_s$ implies the amplitude of the primordial scalar power spectrum, and $n_s$ is the scalar spectral index. To obtain well-behaved sampling as described in \citet{Planck:2013pxb} [with helpful formulae from \citep{Hu:1995en}], the parameter $\theta_{\rm MC}$ is varied in CosmoMC. Then a bisection root finding method is used to obtain the appropriate $H_{0}$ value within the $\Lambda$CDM model. Observables are properly computed using our modified version of \textsc{Camb}.

The parameters $\tilde{\phi}_i$ and $\tilde{\chi}_i$ denote the value of the scalar fields at the scale factor $a_i$ taken to be deep inside the radiation-dominated era. The coupling constant $\tilde{\lambda}_{\chi}$ that is used for the coupling constant of the $\chi$ scalar field, is treated as a free parameter in our MCMC analysis, while the parameter $\tilde{\lambda}_{\phi}$ is a derived parameter, as explained earlier.

The CosmoMC package computes the likelihood of cosmological parameters by including observational data from various sources. We include the combination of CMB, SNe Ia, BAO, and Riess et al. (2019) data sets in our work, and so multiplying the separate likelihoods for these data sets, the total likelihood will be $\mathcal{L}\propto e^{-\chi_{{\rm total}}^{2}/2}$, where $\chi_{{\rm total}}^{2}=\chi_{\mathrm{CMB}}^{2}+\chi_{\mathrm{SN}}^{2}+\chi_{\mathrm{BAO}}^{2}+\chi_{\mathrm{Riess2019}}^{2}$ encodes the deviation between the observational and theoretical results. These data sets were chosen because they are included for use with the standard latest public release of the CosmoMC code. In future work, we will apply the Cobaya \citep{Torrado:2020dgo} or MontePython \citep{Brinckmann:2018cvx} simulation codes to leverage more current data. Following \citet{Poulin:2018cxd, Smith:2019ihp, Poulin:2021bjr, Murgia:2020ryi}, we terminate our MCMC analysis when the Gelman-Rubin convergence criterion \citep{Gelman:1992zz} fulfills $R-1<0.1$. For the statistical analysis of the MCMC chains generated by CosmoMC, we use the publicly available GetDist package \citep{Lewis:2019xzd}, and a burn-in fraction of $0.3$.

To establish priors for CDE (and Rock `n' Roll) parameters in our MCMC as well as an initial guess for best-fit values we begin by finding rough initial guesses for $\tilde{\chi}_{i}$, $\log{(\tilde{\phi}_{i})}$, and $\log{(\tilde{\lambda}_{\chi})}$ that reproduce published Rock `n' Roll \textit{Planck} values for $H_{0}$, the redshift of peak CDE energy-density fraction $z_{c}$, peak CDE energy density-fraction $f(z_{c})$, and $\Omega_{\Lambda}$ when numerically integrated using the equations in Sec.~\ref{section:setup}. We then obtain an initial estimate of best-fit parameters for $\Lambda$CDM + CDE parameters using a simple random-walk simulation. We begin by choosing the standard (but relatively broad) flat priors for the usual cosmological parameters, centered around Planck 2018 best-fit values \citep{Planck:2018vyg}, as well as a trial range for CDE parameters: $\Omega_{b}h^{2}\in\left[0.021,0.024\right],$ $\Omega_{c}h^{2}\in \left[0.10,0.15\right]$, $\theta_{\rm MC}\in\left[1.0,1.1\right]$, $\tau\in\left[0.02,0.09\right]$, $n_{s}\in\left[0.94,1.0\right]$, $\ln(10^{10}{A}_{s})\in \left[2.9,3.2\right]$, $\tilde{\chi}_{i}\in\left[0.01,0.8\right]$, and $\tilde{\lambda}_{\chi}\in \left[14.5,15.5\right]$. An initial guess is made for CDE parameters $\log{\tilde{\phi}}_{i}$, $\tilde{\chi}_{i}$, $\log{\tilde{\lambda}}_{\chi}$ as well as $\Lambda$CDM parameters, and used to compute the likelihood for the full data set (see below) computed within CosmoMC.

Afterward, random guesses are made for all $8$ parameters but are kept only if they improve the likelihood of the model, with a maximum of $100$ iterations. The results are insensitive to the values/ranges initially chosen for $\log{\tilde{\phi}}_{i}$ and $\log{\tilde{\lambda}_{\chi}}$, and indicate a preferred value for $\tilde{\chi_{i}}\simeq 0.4-0.5$. We then use the same flat priors on $\Lambda$CDM parameters and final simulation values to initialize a proper likelihood minimization within CosmoMC, whose initial values are used for the subsequent MCMC. A similar procedure was used for the Rock `n' Roll and single-field DE models considered. We verified that the posterior probability distributions for $\log{\tilde{\phi}}_{i}$ and $\log{\tilde{\lambda}_{\chi}}$ are nearly as flat as the priors, justifying their use without loss of generality. The initial field value $\tilde{\chi}_{i}$ is well constrained in our MCMC and is contained in the assumed prior. The priors used are summarized in Table \ref{table:priors}.

\begin{table*}[!ht]
\centering
\caption{The priors used in our MCMC analysis. When parameters do not appear in a model, we indicate this with a hyphen. The format is [lower bound, upper bound], initial guess.}
\scalebox{0.8}{
\begin{tabular}{|c|c |c |c |c |}
\hline 
{Parameter} & {$\Lambda$CDM} & {Single-field DE} & {Rock `n' Roll} & {Two-field CDE}\tabularnewline
\hline 
$\Omega_{b}h^{2}$ & $\left[0.021,0.024\right],0.023$ & $\left[0.021,0.024\right],0.023$  & $\left[0.021,0.024\right],0.023$& $\left[0.021,0.024\right],0.023 $\tabularnewline
$\Omega_{c}h^{2}$ & $\left[0.11,0.13\right],0.12$ & $\left[0.11,0.13\right],0.12$ & $\left[0.11,0.13\right],0.12$ & $\left[0.11,0.13\right],0.12$\tabularnewline
$100\theta_{\rm MC}$ & $\left[1.02,1.06\right],1.04$ & $\left[1.02,1.06\right],1.04$   & $\left[1.02,1.06\right],1.04$ & $\left[1.02,1.06\right],1.04$  \tabularnewline
$\tau$  & $\left[0.02,0.09\right],0.055$ & $\left[0.02,0.09\right],0.055$  & $\left[0.02,0.09\right],0.055$& $\left[0.02,0.09\right],0.055$\tabularnewline
${\rm ln}(10^{10}A_{s})$ & $\left[2.98,3.12 \right],3.05$ & $\left[2.98,3.12 \right],3.05$& $\left[2.98,3.12 \right],3.05$& $\left[2.98,3.12 \right],3.05$ \tabularnewline
$n_{s}$ & $\left[0.95,0.99\right],0.97$ & $\left[0.95,0.99\right],0.97$  & $\left[0.95,0.99\right],0.97$ & $\left[0.95,0.99\right],0.97$ \tabularnewline
$\log_{\rm 10}(\tilde{\phi}_{i})$ & - & $\left[0.5,2.0\right],1.04$  & - & $\left[0.5,2.0\right],1.04$\tabularnewline
$\tilde{\chi}_{i}$  & - &-  & $\left[0.1,1.0\right],0.54$ & $\left[0.10,1.0\right],0.54$ \tabularnewline
$\log_{\rm 10}(\tilde{\lambda}_{\chi})$ & - & -  & $\left[13.0,15.5\right],14.6$ & $\left[13.0,15.5\right],14.6$ \tabularnewline
\hline 
\end{tabular}
}
\label{table:priors}
\end{table*}

We incorporate the Planck 2018 CMB data for temperature and polarization at small (TT,TE,EE) and large (lowl+lowE) angular scales  \citep{Planck:2018vyg, Planck:2019nip}. We additionally take into account the CMB lensing potential power spectrum measured in the multipole range $40\leq\ell\leq400$ \citep{Planck:2018lbu}. The acoustic peaks are affected by the physics of the decoupling epoch, and their locations are sensitive to physical processes occurring between the decoupling epoch and today.

Type Ia supernovae are standardizable candles that have approximately the same absolute magnitude, once corrections for the width of their light curve are applied. Therefore, they are a powerful tool that can be used to probe the expansion history of the Universe. In our MCMC analysis, we use the Pantheon SN sample \citep{Pan-STARRS1:2017jku}, which consists of magnitude measurements for 1048 SNe Ia with redshifts $0.01 < z < 2.3$. In future work, we will use the more recent Pantheon + sample \citep{Scolnic:2021amr}.

The baryonic acoustic oscillation standard ruler provides a measurement of the angular diameter distance as a function of the cosmological redshift. BAO data can be used to constrain dark energy models. The pressure waves arising from cosmological inhomogeneities in the primordial baryon-photon plasma are imprinted on the CMB and the Large-Scale Structure (LSS) of the galaxy density field. The peak appearing in measurements of the large-scale galaxy correlation function is caused by BAOs. In our analysis, we use BAO measurements from the Baryon Oscillation Spectroscopic Survey (BOSS) \citep{BOSS:2016wmc} ($z\simeq 0.15$), the SDSS Main Galaxy Sample \citep{Ross:2014qpa} ($z\simeq 0.15$), and the 6dFGS \citep{Beutler:2012px} ($z\simeq 0.11$).

Finally, we include the Riess et al. (2019) determination of $H_0 = 74.03 \pm 1.42\,{\rm km\,s^{-1}\,Mpc^{-1}}$ \citep{Riess:2019cxk} for the Hubble constant, based on $70$ Cepheid observations in the LMC and observations of nearby Type-Ia supernovae. This determination is an independent constraint of the expansion rate of the local Universe in our computations.


\section{Results}
\label{section:results}

\begin{table*}[!ht]
\centering
\caption{The best-fit values and 68\% CL constraints for the parameters of the investigated models.}
\scalebox{0.8}{
\begin{tabular}{|c|c  c|c  c|c  c|c  c|}
\hline 
\multirow{2}{*}{Parameter} & \multicolumn{2}{c|}{$\Lambda$CDM} & \multicolumn{2}{c|}{Single-field DE} & \multicolumn{2}{c|}{Rock `n' Roll} & \multicolumn{2}{c|}{Two-field CDE}\tabularnewline
 & best fit & 68\% limits & best fit & 68\% limits & best fit & 68\% limits & best fit & 68\% limits\tabularnewline
\hline 
$\Omega_{b}h^{2}$ & $0.02260$ & $0.02251\pm0.00013$ & $0.02253$ & $0.02251\pm0.00013$ & $0.02298$ & $0.02281\pm 0.00017$ & $0.02285$ & $0.02275\pm 0.00018$\tabularnewline
$\Omega_{c}h^{2}$ & $0.11827$ & $0.11849\pm0.00090$ & $0.11869$ & $0.11851\pm0.00087$ & $0.12173$ & $0.1216\pm 0.0017$ & $0.12049$ & $0.1214\pm 0.0017$\tabularnewline
$100\theta_{\rm MC}$ & $1.04116$ & $1.04116\pm0.00029$ & $1.04108$ & $1.04113\pm0.00029$ & $1.03963$ & $1.03958\pm 0.00073$ & $1.0402$ & $1.03978\pm 0.00074$\tabularnewline
$\tau$ & $0.0596$ & $0.0568\pm0.0071$ & $0.0572$ & $0.0566_{-0.0072}^{+0.0064}$ & $0.0550$ & $0.0540\pm 0.0074$ & $0.0570$ & $0.0540\pm 0.0074$\tabularnewline
${\rm ln}(10^{10}A_{s})$ & $3.048$ & $3.046\pm0.014$ & $3.044$ & $3.046_{-0.014}^{+0.013}$ & $3.050$ & $3.047\pm 0.014$ & $3.053$ & $3.047\pm 0.014$\tabularnewline
$n_{s}$ & $0.9686$ & $0.9690\pm0.0037$ & $0.9698$ & $0.9688\pm0.0036$ & $0.9710$ & $0.9697\pm 0.0036$ & $0.9709$ & $0.9691\pm 0.0037$\tabularnewline
$\log(\tilde{\phi}_{i})$ & $-$ & $-$ & $1.82$ & $2.56_{-0.57}^{+1.3}$ & $-$ & $-$ & $1.43$ & $2.40\pm 0.92$\tabularnewline
$\tilde{\chi}_{i}$ & $-$ & $-$ & $-$ & $-$ & $0.517$ & $0.467^{+0.13}_{-0.075}$ & $0.392$ & $0.428^{+0.14}_{-0.082}$\tabularnewline
$\log(\tilde{\lambda}_{\chi})$ & $-$ & $-$ & $-$ & $-$ & $14.507$ & $-$ & $15.428$ & $-$\tabularnewline
\hline 
$H_{0}$ & $68.76$ & $68.60\pm0.41$ & $68.57$ & $68.60\pm0.41$ & $71.21$ & $70.43\pm 0.99$ & $70.23$ & $70.42\pm 0.98$\tabularnewline
$\Omega_{\rm m}$ & $0.2980$ & $0.7003\pm0.0052$ & $0.3004$ & $0.2997\pm0.0051$ & $0.2854$ & $0.2913\pm 0.0065$ & $0.2906$ & $0.2908\pm 0.0065$\tabularnewline
$\Omega_{\rm DE}$ & $0.7020$ & $0.2997\pm0.0052$ & $0.6996$ & $0.7003\pm0.0051$ & $0.7146$ & $0.7087\pm 0.0065$ & $0.7094$ & $0.7092\pm 0.0065$\tabularnewline
$\sigma_{8}$ & $0.8201$ & $0.8208\pm0.0060$ & $0.8210$ & $0.8208\pm0.0058$ & $0.8394$ & $0.837\pm 0.010$ & $0.8351$ & $0.837\pm 0.010$\tabularnewline
$S_{8}$ & $0.8173$ & $0.820\pm0.010$ & $0.8216$ & $0.8204\pm0.0098$ & $0.8188$ & $0.824\pm 0.010$ & $0.8220$ & $0.824\pm 0.010$\tabularnewline
${\rm Age}/{\rm Gyr}$ & $13.730$ & $13.741\pm0.019$ & $13.724$ & $13.739\pm0.020$ & $13.382$ & $13.46\pm 0.12$ & $13.532$ & $13.49\pm 0.12$\tabularnewline
$\log(\tilde{\lambda}_{\phi})$ & $-$ & $-$ & $-6.34$ & $-9.3_{-4.8}^{+4.1}$ & $-$ & $-$ & $-4.80$ & $-8.7\pm 3.7$\tabularnewline
$z_{c}$ & $-$ & $-$ & $-$ & - &$15700$ & $16790^{+3000}_{-4000}$ & $23500$ & $15990\pm 4000$\tabularnewline
$f(z_{c})$ & $-$ & $-$ & $-$ & - &$0.078$ & $0.052\pm0.021$ & $0.047$ & $0.044\pm 0.022$\tabularnewline
\hline 
\end{tabular}
}
\label{table:parameters}
\end{table*}

\begin{table*}[!ht]
\caption{The resulting values of $\chi^2$ for each model and each data set. The $\chi^2$ values corresponding to different CMB measurements including the Planck lensing power spectrum reconstruction ($\chi^2_{\rm lensing}$), baseline high-$\ell$ Planck power spectra ($30\leq\ell\leq2508$) ($\chi^2_{{\rm high}\ell}$), low-$\ell$ Planck temperature ($2\leq\ell\leq29$) ($\chi^2_{\rm lowl}$), and low-$\ell$ HFI EE polarization ($2\leq\ell\leq29$) ($\chi^2_{\rm simall}$), are also presented in the table. The table also includes the values of $\chi^2_{\rm tot} $ and $\Delta \chi^2  = \chi^2_{\rm Model} -\chi^2_{\rm \Lambda CDM}$.}
\centering
\scalebox{0.8}{
\begin{tabular}{|c|c  c|c  c|c  c|c  c|}
\hline 
\multirow{2}{*}{Parameter} & \multicolumn{2}{c|}{$\Lambda$CDM} & \multicolumn{2}{c|}{Single-field DE} & \multicolumn{2}{c|}{Rock `n' Roll} & \multicolumn{2}{c|}{Two-field CDE}\tabularnewline
 & best fit & 68\% limits & best fit & 68\% limits & best fit & 68\% limits & best fit & 68\% limits\tabularnewline
\hline 
$\chi^2_{\rm lensing}$ & $8.68$ & $9.10\pm 0.55$ & $9.10$ & $9.08\pm 0.50$ & $9.25$ & $9.70\pm 0.90$ & $9.15$ & $9.68\pm 0.88$\tabularnewline
$\chi^2_{{\rm high}\ell}$ & $2348.84$ & $2359.6\pm 5.8$ & $2352.99$ & $2359.7\pm 6.0$ & $2350.79$ & $2360.7\pm 5.8$ & $2350.2$ & $2360.4\pm 5.8$\tabularnewline
$\chi^2_{\rm lowl}$ & $23.04$ & $22.91\pm 0.76$ & $22.03$ & $22.93\pm 0.73$ & $23.02$ & $23.07\pm 0.78$ & $22.94$ & $23.20\pm 0.80$\tabularnewline
$\chi^2_{\rm simall}$ & $397.16$ & $397.2\pm 1.9$ & $395.74$ & $397.2\pm 1.8$ & $395.99$ & $396.8\pm 1.5$ & $396.40$ & $396.8\pm 1.5$\tabularnewline
\hline
$\chi_{{\rm CMB}}^{2}$ & $2777.72$ & $2788.9\pm5.9$ & $2777.19$ & $2788.9\pm6.0$ & $2779.06$ & $2790.2\pm 6.0$ & $2778.69$ & $2790.0\pm 6.0$\tabularnewline
$\chi_{{\rm SN}}^{2}$ & $1034.74$ & $1034.792\pm0.083$ & $1034.74$ & $1034.80\pm0.12$ & $1034.87$ & $1034.87\pm 0.12$ & $1034.85$ & $1034.97\pm 0.29$\tabularnewline
$\chi_{{\rm BAO}}^{2}$ & $5.78$ & $5.89\pm0.81$ & $5.44$ & $5.86\pm0.82$ & $9.79$ & $7.9\pm 2.3$ & $7.33$ & $7.8\pm 2.1$\tabularnewline
$\chi_{{\rm Riess2019}}^{2}$ & $13.79$ & $14.7\pm2.2$ & $14.80$ & $14.7\pm2.2$ & $3.95$ & $6.9\pm 3.6$ & $7.15$ & $6.9\pm 3.5$\tabularnewline
\hline 
$\chi_{\mathrm{total}}^{2}$ & $3832.03$ & $-$ & $3832.17$ & $-$ & $3827.67$ & $-$ & $3828.02$ & $-$\tabularnewline
$\Delta\chi^{2}$ & $0.0$ & $-$ & $0.14$ & $-$ & $-4.36$ & $-$ & $-4.01$ & $-$\tabularnewline
\hline 
\end{tabular}
}
\label{table:chi2}
\end{table*}

\begin{figure*}[t]
\begin{center}
\scalebox{0.90}[0.90]{\includegraphics{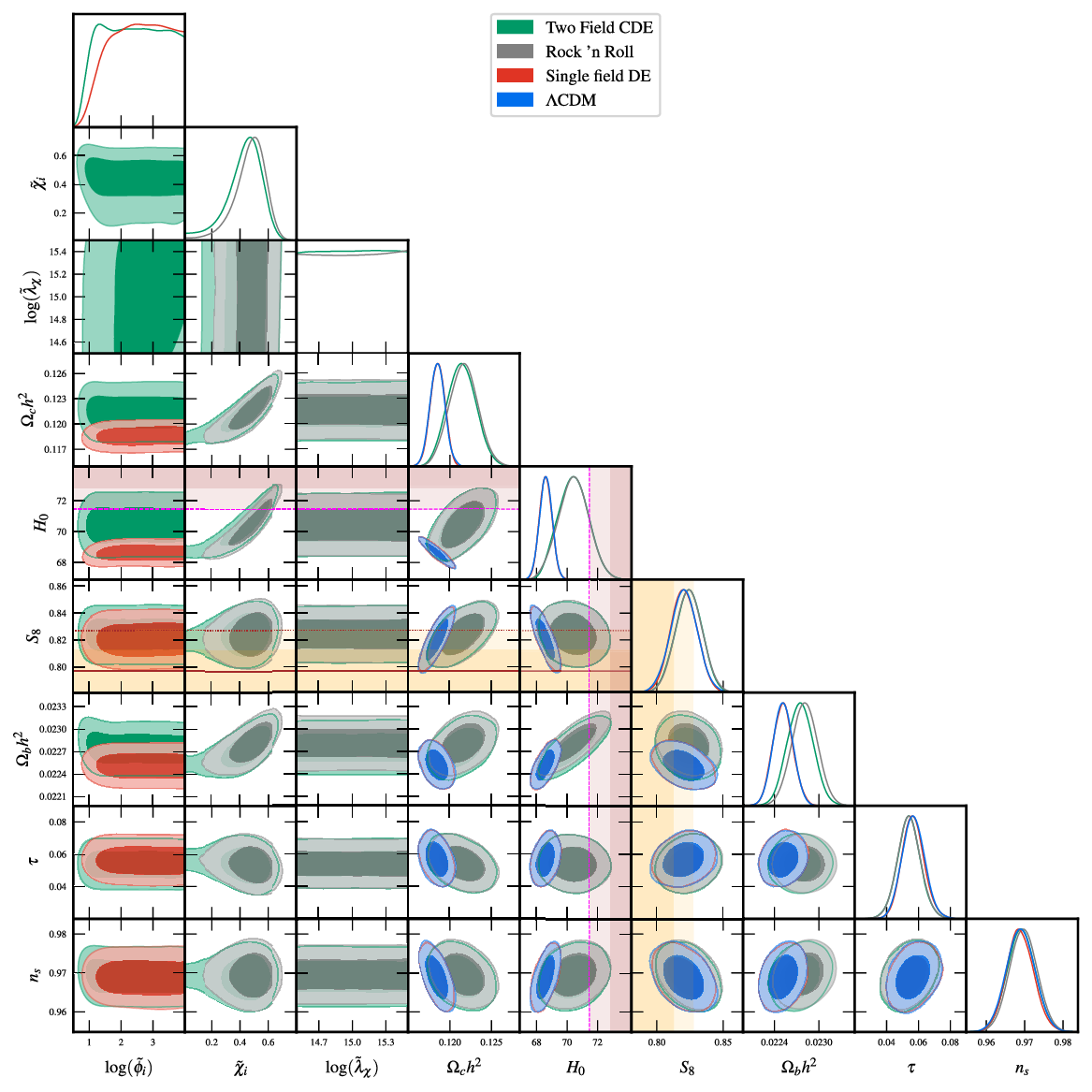}}
\caption{1D likelihoods and 2D contours for the parameters in 68\% and 95\% CL marginalized joint regions for the $\Lambda$CDM model (blue), the single-field DE model (red), the rock `n' roll model (gray), and the two-field CDE model (green). The logarithms for $\tilde{\phi}_i$ and $\tilde{\lambda}_\chi$ are base $10$. The shaded regions in panels where one of the horizontal or vertical axes is $H_{0}$ from \citet{Riess:2019cxk}. The shaded regions in panels where one of the horizontal or vertical axes is $S_{8}$ described in \citet{DES:2021epj}. It should be noted that the $1\sigma$ regions of $S_{8}$ for the $\Lambda$CDM model contours in the figure overlap with the $1\sigma$ bands of DES estimation, so the selected data set is not in tension with the $\Lambda$CDM.}
\label{figure:cde_triangle}
\end{center}
\end{figure*}

\begin{figure*}[t]
\begin{center}
\scalebox{0.75}[0.75]{\includegraphics{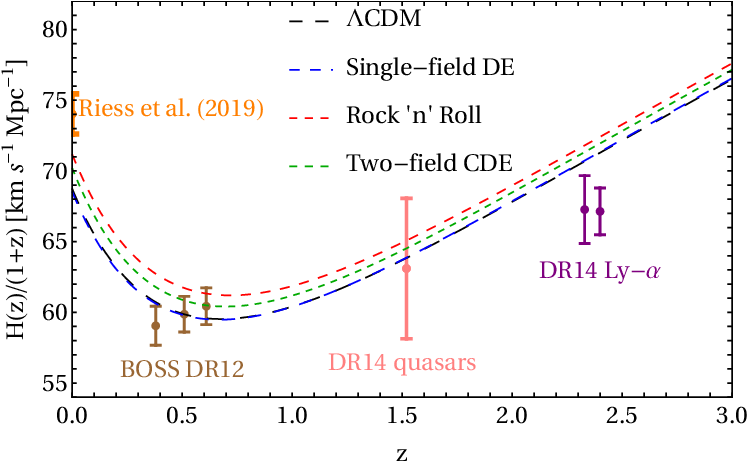}}
\caption{Evolution of the $H(z)/(1+z)$ as a function of cosmological redshift in the $\Lambda$CDM, single-field DE, Rock `n' Roll, and two-field CDE models. Also, in the figure, the data points from \citep{Riess:2019cxk}, BOSS DR12 \citep{BOSS:2016wmc}, DR14 quasars \citep{Zarrouk:2018vwy}, and DR14 Ly-$\alpha$ \citep{Blomqvist:2019rah} measurements have been specified for comparison. Note that only BOSS DR12 data were used to obtain the best-fit model parameters here.}
\label{figure:H}
\end{center}
\end{figure*}

\begin{figure*}[t]
\begin{center}
\scalebox{0.75}[0.75]{\includegraphics{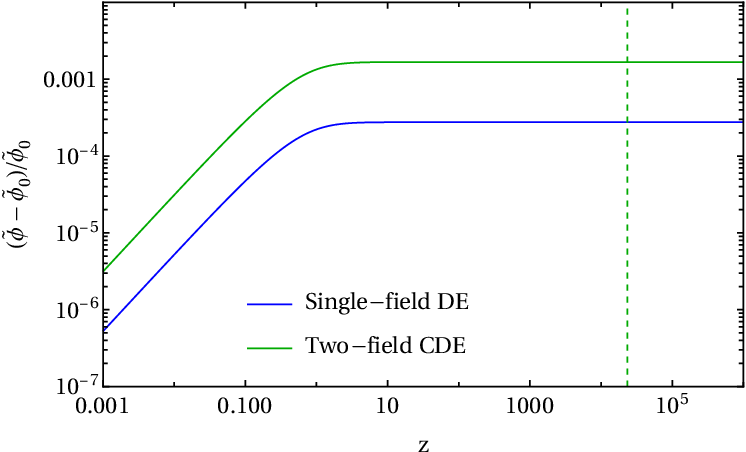}}
\caption{Evolution of the scalar field $\tilde{\phi}$ versus red-shift in the single-field DE and two-field CDE models, with a vertical dashed line indicating $z_{c}$ in the two-field model.}
\label{figure:phi}
\end{center}
\end{figure*}

\begin{figure*}[t]
\begin{center}
\scalebox{0.75}[0.75]{\includegraphics{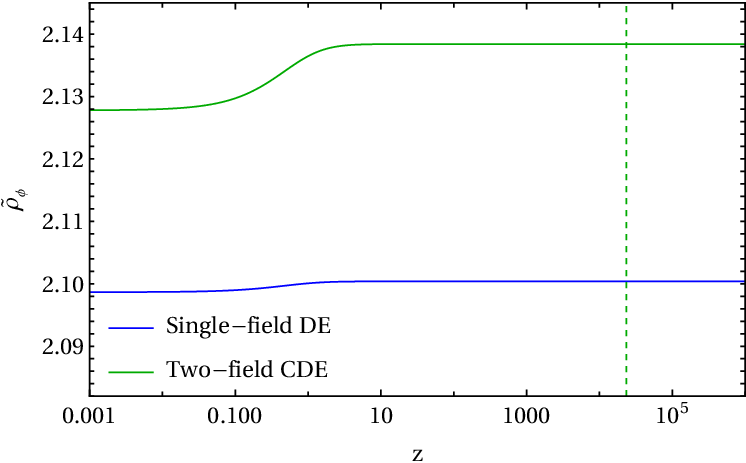}}
\caption{Evolution of the normalized scalar field energy density $\tilde{\rho}_\phi$ as a function of red-shift in the single-field DE and two-field CDE model, with a vertical dashed line indicating $z_{c}$ in the two-field model.}
\label{figure:rho_phi}
\end{center}
\end{figure*}

\begin{figure*}[t]
\begin{center}
\scalebox{0.75}[0.75]{\includegraphics{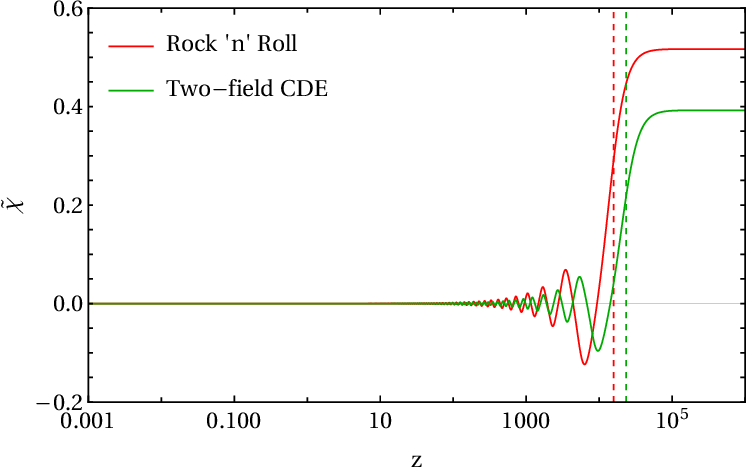}}
\caption{Evolution of the scalar field $\tilde{\chi}$ versus red-shift in the Rock `n' Roll and two-field CDE models, with vertical dashed lines indicating $z_{c}$ in both models.}
\label{figure:chi}
\end{center}
\end{figure*}

\begin{figure*}[t]
\begin{center}
\scalebox{0.75}[0.75]{\includegraphics{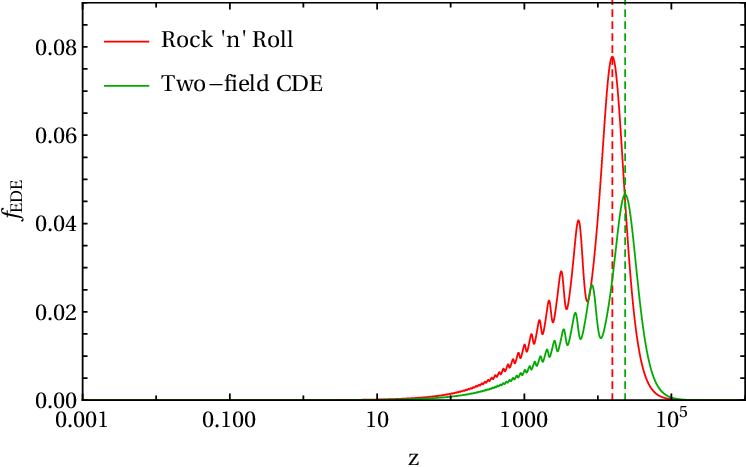}}
\caption{Evolution of the $f_{\rm EDE}$ as a function of cosmological redshift in the Rock `n' Roll and two-field CDE models, with vertical dashed lines indicating $z_{c}$ in both models.}
\label{figure:fEDE}
\end{center}
\end{figure*}

\begin{figure}[t]
\begin{center}
\scalebox{0.85}[0.85]{\includegraphics{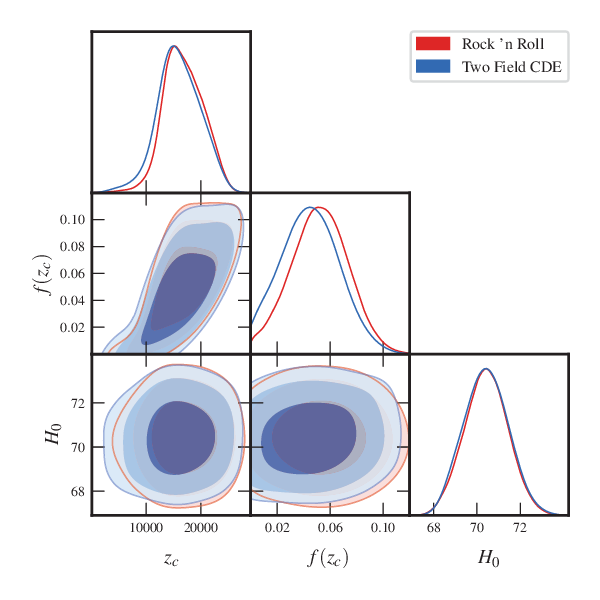}}
\caption{68\% and 95\% CL marginalized 1D likelihoods and 2D contours of $z_c$, $f(z_c)$, and $H_0$ for the Rock `n' Roll model (red), and the two-field CDE model (blue).}
\label{figure:zcfig}
\end{center}
\end{figure}

\begin{figure*}[t]
\begin{center}
\scalebox{0.75}[0.75]{\includegraphics{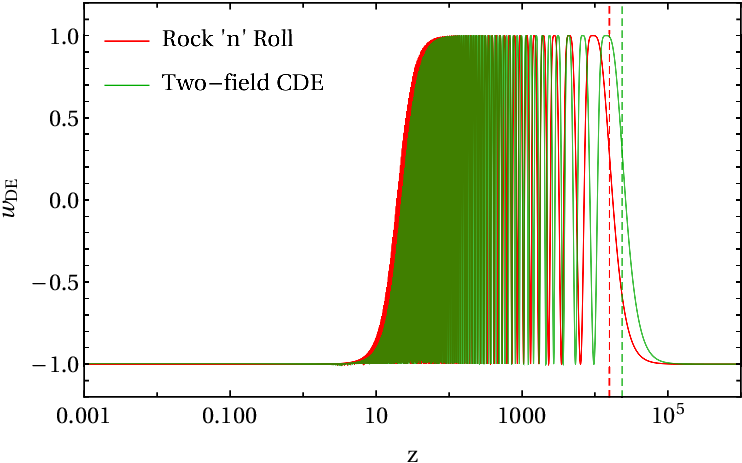}}
\caption{Evolution of the $w_{\rm DE}$ as a function of cosmological redshift in the Rock `n' Roll and two-field CDE models.}
\label{figure:wDE}
\end{center}
\end{figure*}

\begin{figure*}[t]
\begin{center}
\scalebox{0.5}[0.5]{\includegraphics{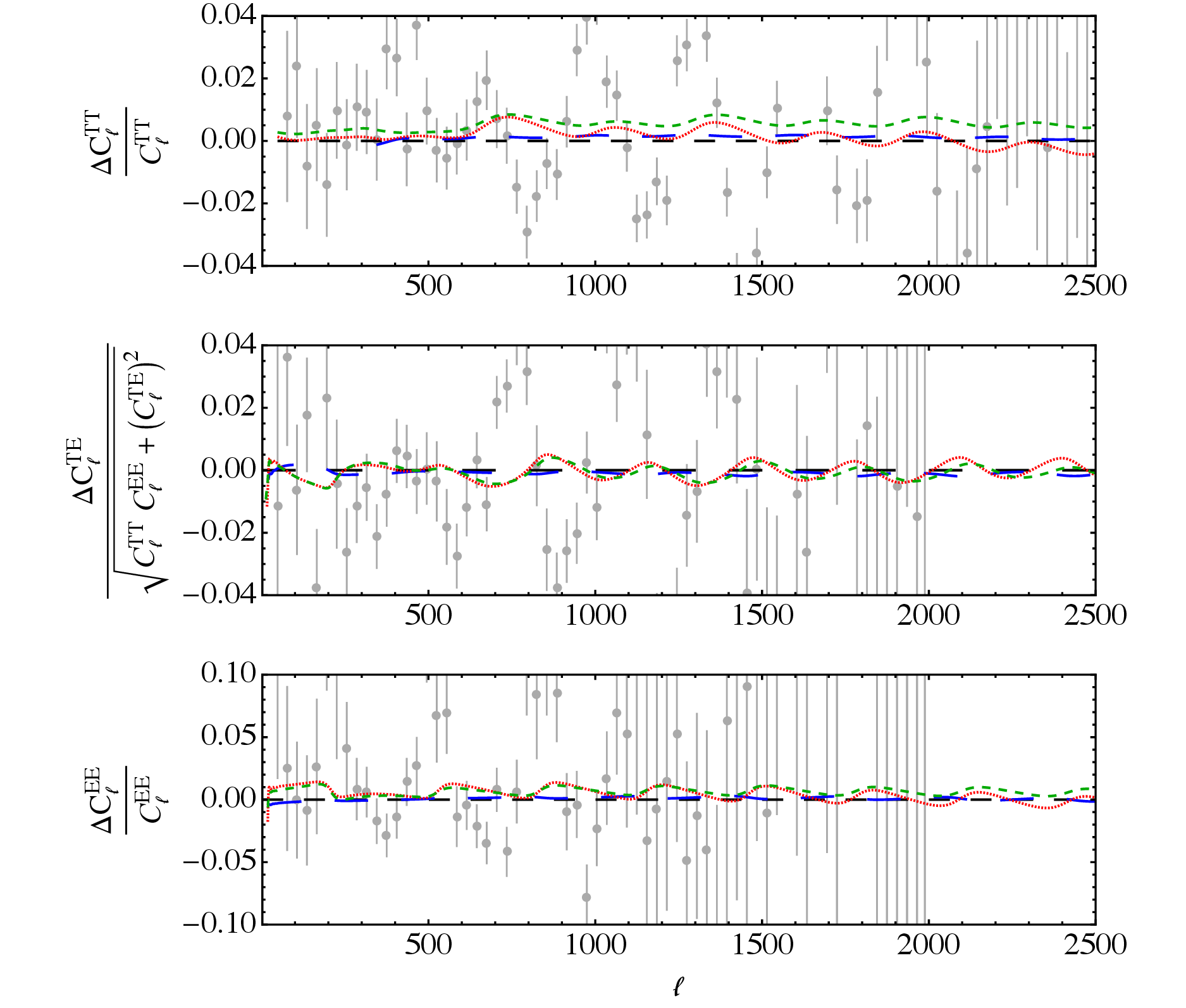}}
\caption{Residuals of the binned CMB power spectra relative to the reference $\Lambda$CDM model. The results of the $\Lambda$CDM (dashed black), single-field DE (long-dashed blue), Rock `n' Roll (dotted red), and two-field CDE (short-dashed green) are compared with the binned Planck 2018 data points (gray) \citep{Planck:2018vyg, Planck:2019nip, Planck:2018lbu}. We compared models using the best-fit parameters for all cases.}
\label{figure:Cls_residuals}
\end{center}
\end{figure*}

The resulting best-fit values and 68\% confidence limit (CL) constraints for the parameters of the investigated models are shown in Table \ref{table:parameters}. In the table, we see that the results for our single-field DE model are less than $0.5\sigma$ away from the $\Lambda$CDM, and the results of the two-field CDE model are much less than $0.5\sigma$ different from those of the Rock `n' Roll scenario. We also see that both the $\Lambda$CDM and single-field DE models return the 68\% CL constraint for the present-day Hubble parameter as $H_{0}=68.60\pm0.41~\mathrm{km\ s^{-1}Mpc^{-1}}$, which disagrees with the value $H_{0}=74.03\pm1.42~\mathrm{km\ s^{-1}Mpc^{-1}}$ measured by Riess et al. (2019) by more than $3\sigma$ \citep{Riess:2019cxk}. We also note that there is a degeneracy between the Rock `n' Roll/CDE parameter $\overline{\chi}_{i}$ and the baryon density parameter $\Omega_{b}h^{2}$, but no degeneracy between $\Omega_{b}h^{2}$ and $\log{(\tilde{\lambda}_{\chi})}$. The other standard $\Lambda$CDM parameters (e.g. $n_{s}$ and $\tau$, the optical depth to reionization) are not noticeably degenerate with Rock `n' Roll/CDE model parameters.

The Rock `n' Roll and two-field CDE models yield a best-fit value for $H_0$ of $H_{0}=70.43\pm 0.99~\mathrm{km\ s^{-1}Mpc^{-1}}$ and $H_{0}=70.42\pm 0.98~\mathrm{km\ s^{-1}Mpc^{-1}}$, respectively. These results are within less than $0.5\sigma$ from one another and also in much better agreement with the Riess et al. (2019) measurement than models with no EDE, reducing the $H_0$ tension to $2\sigma$. The 2D contour plots of the two-field CDE and Rock `n' Roll models which include the $H_0$ parameter are separated thoroughly from the contour plots of the $\Lambda$CDM and single-field DE models.

We see in Table \ref{table:parameters} that the Rock `n' Roll ($S_{8}=0.824\pm0.010$) and two-field CDE ($S_{8}=0.824\pm0.010$) scenarios yield higher values for the $S_8$ parameter by about $0.4 \sigma$ compared with the $\Lambda$CDM ($S_{8}=0.820\pm0.010$) and single-field DE ($S_{8}=0.8204\pm0.0098$) frameworks, and hence their results show more deviations from the DES 3-year observations \citep{DES:2021epj}, which indicate that $S_{8}=0.797_{-0.013}^{+0.015}$. This is a common feature of EDE models \citep{Poulin:2018cxd, Smith:2019ihp, Karwal:2021vpk, Poulin:2021bjr, Murgia:2020ryi}, although it is not statistically significant.


From Table \ref{table:chi2}, we see that the two-field CDE model yields a lower value for $\chi_{\mathrm{total}}^{2}$ than the $\Lambda$CDM or single-field DE models, but is a worse fit than the Rock `n' Roll model ($\Delta \chi_{\rm{total}}^2=0.35$). The table indicates that $\chi_{\mathrm{total}}^{2}$ in the two-field CDE model gets reduced relative to $\Lambda$CDM models and single-field DE by $4.36$ and $4.5$ respectively, but its value has increased by $0.35$ compared to the Rock `n' Roll setup. Although the $\chi_{\mathrm{total}}^{2}$ value in the Rock `n' Roll scenario is less than the two-field CDE model ($\Delta\chi_{\rm{total}}^{2}=0.35$), it is interesting to point out that the two-field CDE model is a better fit to the BAO data in comparison with the Rock `n' Roll model ($\Delta\chi_{\mathrm{BAO}}^{2}=-2.46$) which implies that BAO data favor late-time evolution in the DE component.

Results for 1D likelihoods and 2D contour plots in the 68\% and 95\% CL regions are shown in Fig. \ref{figure:cde_triangle}. The graph indicates that the results of the single-field DE model are less than $0.5\sigma $ of those of the $\Lambda$CDM model, and also that the results of the two-field CDE model are within a $0.5\sigma$ interval of the Rock `n' Roll results.


The value of $\tau$, the optical depth parameter, decreases in the two-field CDE model, compared to the $\Lambda$CDM and single field DE scenario. In the Rock `n' Roll model, the value of this parameter gets reduced relative to the two field CDE model by $1\sigma$. Compared with \citet{Agrawal:2019lmo}, our best-fit RnR values for $A_{s}$ and $n_{s}$ are further enhanced compared with $\Lambda$CDM.

These differences are likely due to the neglect of perturbations, and may be responsible for the fact that our best-fit RnR $\Delta \chi^{2}$ is $-4.4$, rather than the $-7.8$ reported in \citet{Agrawal:2019lmo}. An important factor in this change is worse agreement between our data and the SH0ES data set, when marginalized over model parameters, which have been pushed into different regions of the posterior volume due to details of perturbations. Since the statistically preferred parameter region for $\phi$ is such that the field has not yet started coherently oscillating, the perturbations in $\phi$ should be relatively small and $\phi$ is approximately unclustered. We thus anticipate that the relative difference in $\Delta \chi^{2}$ between RnR and CDE is stable to the inclusion of perturbations. In any case, our work thus provides a specific model realization of the two-fluid scenario also explored using a fluid parameterization in \cite{Moshafi:2022mva}.

Figure~\ref{figure:cde_triangle} additionally shows that the likelihood is quite insensitive to $\tilde{\phi}_i$ and $\tilde{\lambda}_{\chi}$, as evidenced by the flatness of the posterior contours in these parameters. These facts merit some explanation. For any $\phi_i$ bigger than a threshold, a larger value can be compensated with a smaller value of $\log \lambda_{\phi}$ where $\lambda_{\phi}\phi_i^4 $ is of order the observed cosmological constant. Regarding $\chi$, we can apply the approximate expression for $f_{\rm EDE}$ from \cite{Smith:2019ihp} in the anharmonic limit for $n=2$ (corresponding to a $\lambda_{i} \phi_{i}^{4}$ potential), to find that $f_{\rm EDE}\sim \chi_{i}^2/M_{\rm pl}^2$. The coupling thus drops out and is roughly unconstrained, while $\chi_{i}$ is related to $f_{\rm EDE}$ leading to a meaningful constraint.

There is a preferred value for $\tilde{\chi}_i$ in the two-field CDE and Rock `n' Roll models, implying a non-vanishing contribution of EDE to the cosmic energy density.

In Fig.~\ref{figure:H}, we used the best-fit values of the parameters listed in Table \ref{table:parameters} to plot the evolution of the Hubble parameter against redshift, showing that the Hubble parameter of the single-field DE model is coincident with $\Lambda$CDM. Also, the results of the Rock `n' Roll and two-field CDE models are $\sim 1\sigma$ away from each other. Nevertheless, at both the earlier and late times, the Hubble parameter of the Rock `n' Roll scenario exceeds the value in the two-field CDE model, yielding a greater $H_0$.

Smaller values of the Hubble parameter at lower redshifts give better compatibility of the two-field CDE model with BAO data, compared to Rock `n' Roll ($\Delta\chi_{\mathrm{BAO}}^{2}=-2.46$). Simultaneous agreement with PANTHEON and BAO data causes the $\phi$ field to be over-constrained by the data, limiting the ability of the CDE model to better fit BAO data. It may be possible to go beyond this limitation by adding additional scalar fields. We have also developed a modified version of \textsc{Camb} that has \textbf{three} dynamical fields ($\chi$, $\phi$, and a field $\xi$ that rolls at an intermediate redshift). Preliminary exploration of some parameter combinations leads us to speculate that this scenario could provide a significantly better fit to lower-$z$ BAO data (e.g. as in Fig. \ref{figure:H}), but a proper MCMC analysis of this scenario using cosmological data awaits future work. As the $3^{\rm}$ field-model is likely to begin coherent oscillation between recombination and the present-day dominance of dark energy, perturbations will become essential in obtaining the \emph{relative} improvement of 3-field models over RnR and CDE. In pursuit of that scenario, we will compute the full perturbative dynamics of the $3$-field scenario, taking advantage of the more robust \textsc{CLASS} and Cobaya codes to ensure that this task is completed in a computationally efficient manner.

In Fig.~\ref{figure:phi}, we show the evolution of the normalized scalar field $\tilde{\phi}$ in the single-field DE and two-field CDE models as a function of the scale factor. The accompanying scalar-field energy tensity $\tilde{\rho}_{\phi}$ is shown in Fig.~\ref{figure:rho_phi}. Aside from a small interval at late times, the scalar field $\tilde{\phi}$ tends to remain constant. But, around the present epoch, it becomes dynamic. The fractional variation of $\tilde{\phi}$ in the (best-fitting) single-field scenario is substantially greater than in the (best-fitting) two-field CDE scenario - this is because simultaneously allowing late and early-time dark energy allows the model to fit both the late-time acceleration of the Universe and provide the required early-time reduction in the sound horizon needed to resolve the Hubble tension. In some sense, demanding less of the $\phi$ field allows its late-time behavior to more closely resemble $\Lambda$CDM.

In Fig.~\ref{figure:chi}, the evolution of the scalar field $\tilde{\chi}$ as a function of scale factor is plotted for the Rock `n' Roll and two-field CDE setups. Here we have used the best-fit values given in Table \ref{table:parameters}. The figure shows that the scalar field $\tilde{\chi}$ is almost constant at earlier times in these models, and after some time, it begins to oscillate around the potential minimum at $\tilde{\chi} = 0$. The initial value of $\tilde{\chi}$ is smaller in the two-field CDE model compared to the Rock `n' Roll model, and consequently, its oscillations occur in this model earlier than in the Rock `n' Roll model. At late times, the amplitude of the $\tilde{\chi}$ oscillations becomes very small, and accordingly, its contribution to the matter-energy content of the Universe becomes negligible.

In Fig.~\ref{figure:fEDE}, we show the fraction of the EDE energy density relative to the total energy density, that is, $f_{\rm EDE}\equiv\rho_{\chi}/\rho_{\mathrm{total}}$. We see that $f_{\rm EDE}$ is negligible at the initial times, but after a while, it grows sharply and reaches a peak at the critical redshift $z_c$, and then it drops again and becomes very small at the late-times. For the Rock `n' Roll model, the peak appears at the critical redshift $z_{c}=1.57\times10^{4}$ with the maximum value $f_{\rm EDE}=0.078$. The peak of the two-field CDE model appears at $z_{c}=2.35\times10^{4}$ with the maximum amplitude $f_{\rm EDE}=0.047$. The earlier peak energy of the two-field CDE model means that this variant of EDE has less of an impact on the sound horizon at recombination \cite{Knox:2019rjx}, and in turn explains the result of our MCMC analysis, which shows that the two-field model is $\sim 1\sigma$ worse in relieving the Hubble tension than the Rock `n' Roll model.

In prior work on EDE models, the parameters $z_{c}$ and $f_{\rm EDE}$ [rather than model-specific parameters such as $\log{(\tilde{\phi_{i}})}$, $\tilde{\chi}_{i}$, or $\log{(\tilde{\lambda}_{\chi})}$] are sometimes varied in the MCMC. We output these parameters as derived values from CosmoMC (with an appropriate modification to \textsc{camb}), and show the resulting posteriors in Fig.~\ref{figure:zcfig} and Table~\ref{table:parameters}. We see that there is a strong degeneracy between $f(z_{c})$ and $z_{c}$ in both EDE models considered. We verified that the data yield a significantly smaller allowed region than the effective prior for these derived parameters. The difference between the best fit values $z_{c}$ [or $f(z_{c})$] obtained from likelihood and posterior in Fig.~\ref{figure:fEDE} results from the highly non-Gaussian posterior and range of $f(z_{c})$ (or $z_{c}$) values which must be marginalized over.

Figure~\ref{figure:wDE} shows the evolution of the equation of state parameter of dark energy, $w_{\rm DE}\equiv p_{\rm DE}/\rho_{\rm DE}=\left(p_{\phi}+p_{\chi}\right)/\left(\rho_{\phi}+\rho_{\chi}\right)$, in terms of cosmological redshift. At high redshifts, $w_{\rm DE}$ begins at $-1$, and subsequently oscillates around zero, with $w_{\rm DE}$ varying between $-1$ and $1$. Eventually, it converges to $-1$ and remains very close to this value until the present time. The oscillations of $w_{\rm DE}$ start sooner in the two-field CDE model than in the Rock 'n Roll scenario, although the oscillations last longer for Rock 'n Roll.

We present the residuals of CMB power spectra in Fig.~\ref{figure:Cls_residuals}. In the figure, we have also shown Planck 2018 data points \citep{Planck:2018vyg, Planck:2019nip, Planck:2018lbu} for comparison. We see in the figure that the results of the single-field DE are indistinguishable from $\Lambda$CDM results, while the Rock `n' Roll and two-field CDE models show deviations from $\Lambda$CDM, exceeding cosmic variance in some cases. To roughly estimate the improvement in constraining power of future CMB experiments, we compute the statistic \citep{Grin:2009ik}
\begin{equation}
    \mathcal{Z}\equiv \sqrt{\sum_{\ell}\frac{\Delta C_{\ell}^{2}}{\sigma_{\ell}^{2}}},
\end{equation}
where the sum over the binned spectrum is evaluated using the deviation $\Delta C_{\ell}$ of the model of interest, with binned cosmic variance errors used for $\sigma_{\ell}$. We separately compute this sum for temperature and $EE$ polarization. A full forecast requires a proper analysis of T and E covariances, or better yet, a full Fisher matrix analysis that properly accounts for parameter degeneracies including nuisance parameters). Nonetheless, this rough estimate can give us some sense of whether or not the CMB alone can detect deviations between the models we consider, with future data. Roughly speaking, $\mathcal{Z}$ is the number of sigmas at which two scenarios can be distinguished.

Using temperature, we find that future data could distinguish between the single and two-field scenarios with $\sim 3\sigma$ significance, which improves to a $\sim 20\sigma$ potential detection in the limit that TE covariances vanish. Of course, this is not the case, and so a full Fisher matrix forecast is likely to show an answer closer to $\sim 7\sigma$ (ie, the geometric mean of the two extreme cases).


\section{Conclusions}
\label{section:conclusions}
 
We studied the cascading dark energy model, in which many scalar fields contribute to the dark energy component of the Universe. In this setup, a large number of canonical scalar fields with sub-Planckian field excursions and steep potentials cooperate to provide a super-Planckian excursion and a relatively flat potential that can induce the late acceleration of the Universe as well as an early dark energy epoch. In the example we considered, the discordant initial conditions between fields cause some to cascade, drop out of the ensemble, and start oscillating around their minima. We restricted our attention to a single cascade involving an interplay reducing to an effective theory of two fields, $\phi$, and $\chi$. We choose the initial conditions such that $\phi$ plays the role of the late dark energy, while the $\chi$ field behaves as EDE, subsequently cascading and decaying quickly.

We used MCMC simulations to test  CDE and related scenarios using CMB, BAO, PANTHEON observations of Type-Ia supernovae, and Riess et al.~(2019) data. Our work shows that in the CDE model, $\chi_{\mathrm{total}}^{2}$ gets reduced relative to $\Lambda$CDM and single-field DE models by $4.01$ and $4.15$ respectively. In contrast, the CDE model's $\chi_{\mathrm{total}}^{2}$ is worse than the Rock `n' Roll model with  $\Delta \chi_{\mathrm{total}}^{2}=0.35$. The two-field CDE model yields a fit of $H_0 = 70.42\pm 0.98~\mathrm{km\ s^{-1}Mpc^{-1}}$, substantially higher than $\Lambda$CDM ($H_0 = 68.60\pm0.41~\mathrm{km\ s^{-1}Mpc^{-1}}$) and single-field DE ($H_0 = 68.60\pm0.41~\mathrm{km\ s^{-1}Mpc^{-1}}$) values, while very close to the Rock `n' Roll values ($H_0 = 70.43\pm 0.99~\mathrm{km\ s^{-1}Mpc^{-1}}$).

The CDE and Rock `n' Roll versions of the EDE scenario thus reduce the Hubble tension between supernovae ($H_{0}=74.03\pm1.42~\mathrm{km\ s^{-1}Mpc^{-1}}$) and other methods. Our analysis also shows that the two-field CDE model ($\chi_{{\rm BAO}}^{2} = 7.33$) provides a modestly improved fit to BAO data compared with the Rock `n' Roll model ($\chi_{{\rm BAO}}^{2} = 9.79$). Of course this comes at the price of worse compatibility with the Riess $H_{0}$ data, which makes the overall $\chi^2$ worse than the Rock `n' Roll model.

There are a number of important avenues needed to expand and critically test our results. The analysis shown here did not follow the linear perturbations of either DE component - as noted, these do have a statistical impact on inferences about EDE properties, and in the future, we will generalize our analysis to include CDE clustering in the linear regime \citep{inp3fpert}. We will also explore the possibility of rich resonant non-linear phenomenology, as discussed in \citet{Smith:2019ihp}. Another area for future work is the exploration of a broader range of potential energy functions, motivated by a variety of considerations.

Since the effective super-Planckian field samples the potential at relatively large values of the scalar field, the form of the potential at such values of the field can effectively be different from the part of the potential that only samples the sub-Planckian values of the field. For example, the late field can effectively be on the parts of the potential that is just a cosmological constant. In this limit, the predictions and statistical significance of the model should be comparable approaches to the Rock `n' Roll model. However, in principle, the late and early dark energy fields can have different potentials and this affects the predictions of the model. This is one interesting research avenue that is worth considering in the future.

Even with monomial potentials for both the late and early dark energy fields, one can assume non-renormalizable forms for the potentials. For example with a sixth-order monomial potential, $\phi^6$, we expect that the model can achieve larger values of $H_0$, as shown in \cite{Poulin:2018cxd}. The scalar fields also may have non-canonical kinetic terms. In particular, we can assume the kinetic terms for the dark energy fields have a DBI form, which has well-based theoretical motivations \citep{Ahn:2009xd}. In addition, although in this paper, we assumed that the late dark energy and the cascading fields are coupled to the Einstein gravity minimally, one can consider their non-minimal couplings with gravity too \citep{Sakstein:2019fmf, Braglia:2020auw, Karwal:2021vpk, McDonough:2021pdg}. These possibilities are left for future investigation.


\acknowledgments
D.~G. acknowledges support in part by NASA ATP Grant No.~17-ATP17-0162, and the provost's office of Haverford College. We thank Tristan Smith for useful conversations. This project has received funding/support from the European Union’s Horizon 2020 research and innovation programme under the Marie Sklodowska-Curie grant agreement No 860881-HIDDeN.



\bibliography{bibliography}{}
\bibliographystyle{aasjournal}



\end{document}